# Interaction between Metamaterial Resonators and Inter-subband transitions in Semiconductor Quantum Wells

Alon Gabbay,[1,*] John Reno[1], Joel R. Wendt[2], Aaron Gin[1,2], Michael C. Wanke[2], Michael B. Sinclair[2], Eric Shaner[1] and Igal Brener[1,2]

[1]*Center for Integrated Nanotechnologies, Sandia National Laboratories, P.O. Box 5800, Albuquerque, New Mexico 87185, USA*

[2]*Sandia National Laboratories, P.O. Box 5800, Albuquerque, New Mexico 87185, USA*

[*]*Corresponding author: angabba@sandia.gov*

We report on the coupling and interaction between the fundamental resonances of planar metamaterials (split ring resonators) and inter-subband transitions in GaAs/AlGaAs quantum wells structures in the mid-infrared. An incident field polarized parallel to the sample surface is converted by the metamaterial resonators into a field with a finite component polarized normal to the surface, and interacts strongly with the large dipole moment associated with quantum well inter-subband transitions.

Metamaterials have been the subject of intense research in the past few years. They give rise to new capabilities for the manipulation of electromagnetic radiation, such as left handed refraction, non-diffraction limited imaging and cloaking [1, 2]. They can also be used to probe strong and weak coupling using a single sheet of material [3, 4]. The realization of tunable metamaterials is emerging as a natural subtopic in this field. They can be used as optical switches, optical modulators and phase shifters at wavelengths where these types of devices do not exist.

Recently, several approaches have been attempted in order to realize tunable metamaterials at infrared (IR) and terahertz (THz) frequencies. Amplitude and phase modulation were achieved at THz frequencies by depletion of carriers in doped GaAs layers[5, 6]. At higher frequencies, (mid-IR) frequency tuning of metamaterial response was achieved by using InSb epilayers with different doping levels [7]. In the near-IR range frequency tunability was demonstrated by thermally/electrically induced insulator-to-metal phase transition in vanadium dioxide ($VO_2$) [8].

In this paper, we propose and provide evidence on the interaction between metamaterial resonators and intersubband transitions in semiconductor heterostructures. We utilize inter-subband transitions (IST) in semiconductor quantum wells (QWs) to control the dielectric response of the sample on which split ring resonators (SRRs) are fabricated. The major advantage of ISTs is the wide scalability in wavelength response that can be obtained through QW structural parameters such as the doping level, energy spacing between subbands and the use of different material systems. This scalability can be combined with the scalability of planar metamaterials in order to provide tunable devices that could span the near to far IR[9, 10, 11].

Metallic split ring resonators are one of the most studied metamaterial resonators [12, 13]. The fundamental mode of the SRR can be excited in a TE geometry or at normal incidence (in both cases the incident electric field is polarized parallel to the gap in the y-direction, as defined in Fig. 1(a)) [14]. For this polarization and at resonance, most of the electromagnetic energy is concentrated in the gap region. Although the strongest electric field component in the gap region occurs in the y direction, there is also a component normal to the sample (E- field pointing in the z direction). We used FDTD Solutions [15] to model the electromagnetic response of an array of SRRs together with the anisotropic response of the semiconductor heterostructures directly underneath. In Fig. 1(b) we show the calculated transmission spectra through the SRR array on a QW sample and compare it to an array of closed square loop rings of identical dimensions. The latter serves as a control sample where the lowest metamaterial resonance is absent.

We then proceed to calculate field cross sections for excitation at the resonant frequency, denoted by the arrow in Fig. 1(b). Fig 1(c) shows a cross section in the plane parallel to the y-z axes, that is located at the center of the gap in the x direction. The maximum field intensity is at the interface of the metal and the sample and at the edge of the SRR, with the field amplitude decaying as the distance from the SRR increases towards the substrate. Fig 1(d) shows a cross section of the field along the x-y plane 15nm from the interface in the substrate. At this depth, there is a field enhancement at the edge of the gap region of ~2.5 times the incident field. Fig 1(e) shows similar field plots at a plane located 65 nm away from the interface. The maximum z-component of the field at this depth is about 1.5 times stronger than the y-component of the input field amplitude. Figs. 1 (f,g) show field profiles calculated at the same frequency and the same depths of 15nm and 65 nm as in figure 1(d,e), but for the control case, namely a closed square loop. The closed square loop has no resonance at this frequency and there is no field enhancement. Clearly, the SRRs do not act simply as a subwavelength grating [16] but concentrate the optical field in the gap region. Some of this enhanced field has significant amplitude in the z direction and thus can couple directly to the IST dipole of the QWs.

From our modeling and analysis we conclude that there are three parameters which need to be maximized in order to increase the coupling strength between the SRRs and the ISTs. The first parameter is electron density (Eq. 1). The second parameter is the intersubband matrix element, $z_{21}$, which can be optimized using bandgap engineering [9]. The third parameter is the proximity of the QWs to the metallic traces of the SRR. The closer the QWs are to the metal traces the stronger the coupling will be.

In this work we used two coupled QWs as the basic unit cell (Fig. 2(b)). The quantum well structure was grown by molecular beam epitaxy (MBE) and consists of the following: The basic unit cell is a coupled QW structure **15**/5.75/**1.13**/2.5/**15** nm thick layers of (**Al$_{0.5}$Ga$_{0.5}$As**/GaAs), followed by a Si $\delta$-doped layer (Fig. 2(a)). The unit cell repeats itself 30 times and is grown on top a semi-insulating GaAs substrate. The electron density within each coupled QW structure is about $2\times10^{11}$ cm$^{-2}$.

A number of split ring resonator arrays were fabricated using electron beam lithography, metal deposition and lift-off. The basic pattern has Ti/Au metal traces 10/70 nm thick, a period of $1.14\,\mu m$ and is repeated to form 3x3 mm$^2$ arrays. The dimensions of a few representative SRR structures are shown in Fig. 2(c). In order to vary the resonance frequency around the IST resonance frequency, several arrays were fabricated where all dimensions were linearly scaled[10] (length, linewidth, gap size and periodicity) except for the metal trace thickness which was kept uniform. In order to provide additional control samples for distinguishing the effects of the SRR and the IST, the same SRR and square loop arrays were fabricated both on the QW sample and on a SI- GaAs wafer.

Polarized transmission measurements as a function of frequency were measured using an FTIR (Bruker Vertex 80v). Representative spectra through 3 different SRR arrays are shown in Fig. 3(a-c), and the basic pattern is scaled by 0.9, 1.0 and 1.1, respectively. The thick solid blue curve in each figure represents the transmission through the SRR array fabricated on top of the QW sample, where the incident light is polarized parallel to the SRR gap (y direction). The thin solid black curve corresponds to incident light polarized along the x direction. The thick red and thin gray dotted lines correspond to transmission spectra measured for SRRs deposited on top of a SI-GaAs sample and with incident light polarized in the y and x directions, respectively. All the spectra are normalized relative to the transmission spectra measured on the corresponding sample with no metamaterials.

Evidence for coupling between the SRR lowest resonance and the ISTs can be inferred from the thick blue solid transmission spectra lineshape. As the SRR resonance overlaps the IST transition energy, its linewidth broadens significantly. This is in contrast to the case of SRRs on a bare SI-GaAs substrate where the broadening of this resonance does not change significantly and is merely dictated by ohmic losses in the metal and radiation losses as the resonant frequency is changed (thick red dotted line).

The transmission spectra of the incident x-polarized field (thin solid black lines of Fig. 3(a-c)) show a feature which has its minimum at 25.5 THz. This feature corresponds to absorption of incident light by the ISTs. For this polarization and frequency, the metamaterial elements act as a sub-wavelength grating. Evidence for that can be seen by comparing the transmission through a closed square loop (thin green dashed curve) to the transmission through the SRR when the incident polarization is in the x-direction (thin solid black lines) in Fig. 3(c). The two transmission lines are almost identical. The grating's fringing fields have a finite component of the electric field in the growth direction [16] and this field is absorbed by the ISTs.

The optical response of electrons occupying the first subband and excited to the second subband in a QW can be modeled using a Lorentzian oscillator:

$$\chi_{21}(E) = -\frac{Ne^2 |z_{21}|^2}{\varepsilon_0} \frac{1}{E-(E_2-E_1)+i\gamma} \quad (1)$$

Where $E_1$ and $E_2$ denote the first and second conduction subband energies. N is the volume electron density in the QW, $\varepsilon_0$ is the vacuum permittivity and $z_{21}$ is the inter-subband matrix element [9].

We use the susceptibility provided by Eq. (1) in the numerical modeling of the coupled metamaterial-QW system. However, using the nomial values given by Eq. (1) provided a coupling too small to replicate the measurements shown in Fig. 3. Good agreement with experiments was obtained when the oscillator strength of the intersubband resonance was increased by 4 and this is shown in Fig. 4, and for both polarizations. A best fit to the experimental data was obtained for $\gamma = 6$ meV, $N = 3.5 \times 10^{17}$ cm$^{-3}$. In the simulations, the interaction between the two resonances (metamaterial and IST) is manifested through line splitting much as in the case of any two coupled resonators. In the experimental data, this level splitting is not completely resolved and it could be indicative of other inhomogeneous broadening mechanisms that act on the intersubband transition. This correction indicates that the experimental interaction between the SRRs and the ISTs is much stronger than what the model predicts (when using the nominal parameter values). A possible reason could be increased scattering of the incident optical field from defects or fabrication imperfections in the metal traces, increasing thus the z component of the field inside the semiconductor.

In conclusion, we have shown experimentally the existence of coupling between SRRs and ISTs in GaAs QWs, at a frequency of about 25.5 THz (wavelength of 11.8 μm). The enhanced field in the gap region at the resonant frequency includes enhanced field amplitude in the z direction, which then couples to the inter-subband transitions. The coupling is manifested through a line broadening in the transmission spectra relative to the transmission spectra through SRRs on a SI-GaAs sample.

We thank James Ginn and Dale Huber for technical assistance. The simulation portion was funded by the Energy Frontier Research Center for Solid State Lighting Science. Other parts were funded by DARPA/MTO's CEE program under DOE/NNSA Contract DE-AC52-06NA25396. This work was performed, in part, at the Center for Integrated Nanotechnologies, a U.S. Department of Energy, Office of Basic Energy Sciences user facility. Sandia National Laboratories is a multi-program laboratory operated by Sandia Corporation, a Lockheed-Martin Company, for the U. S. Department of Energy under Contract No. DE-AC04-94AL85000.

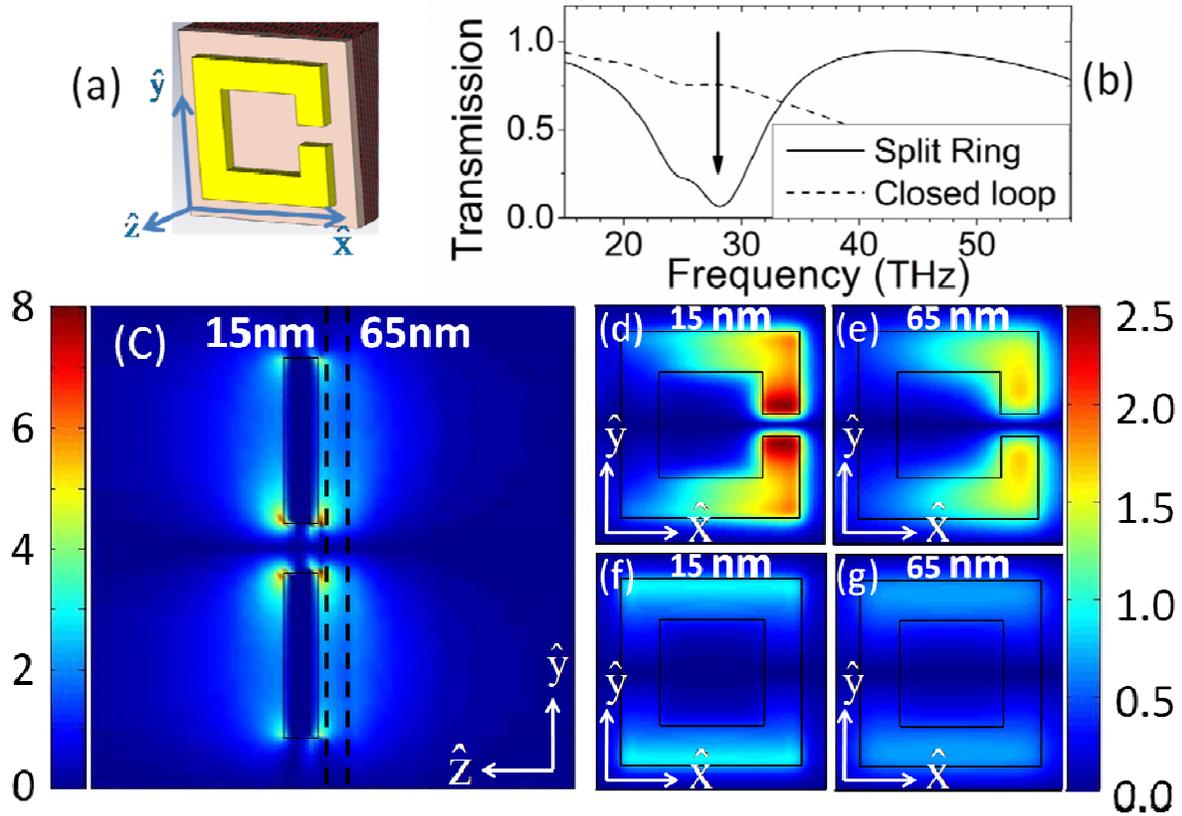

Fig. 1: (a) A schematic diagram of the SRR on top of a QW (the dimensions are shown in Fig. 3(c)), and its calculated transmission spectrum (b). The dashed curve is the calculated transmission through a closed square loop on top of the QW sample. The arrow points at the frequency at which Figs. (c-g) were calculated. (c-g) Cross sections of the z-component of the electric field parallel to the axes indicated on each figure. The color bars indicate the ratio between the shown field and the incident field (polarized in the y-direction). (c) Located in center of the gap in the x-direction. (d,e) Located 15nm and 65nm away from the SRR/sample interface into the sample, respectively. (f,g) Are the same as (d,e) but calculated for the closed square loop instead of a SRR.

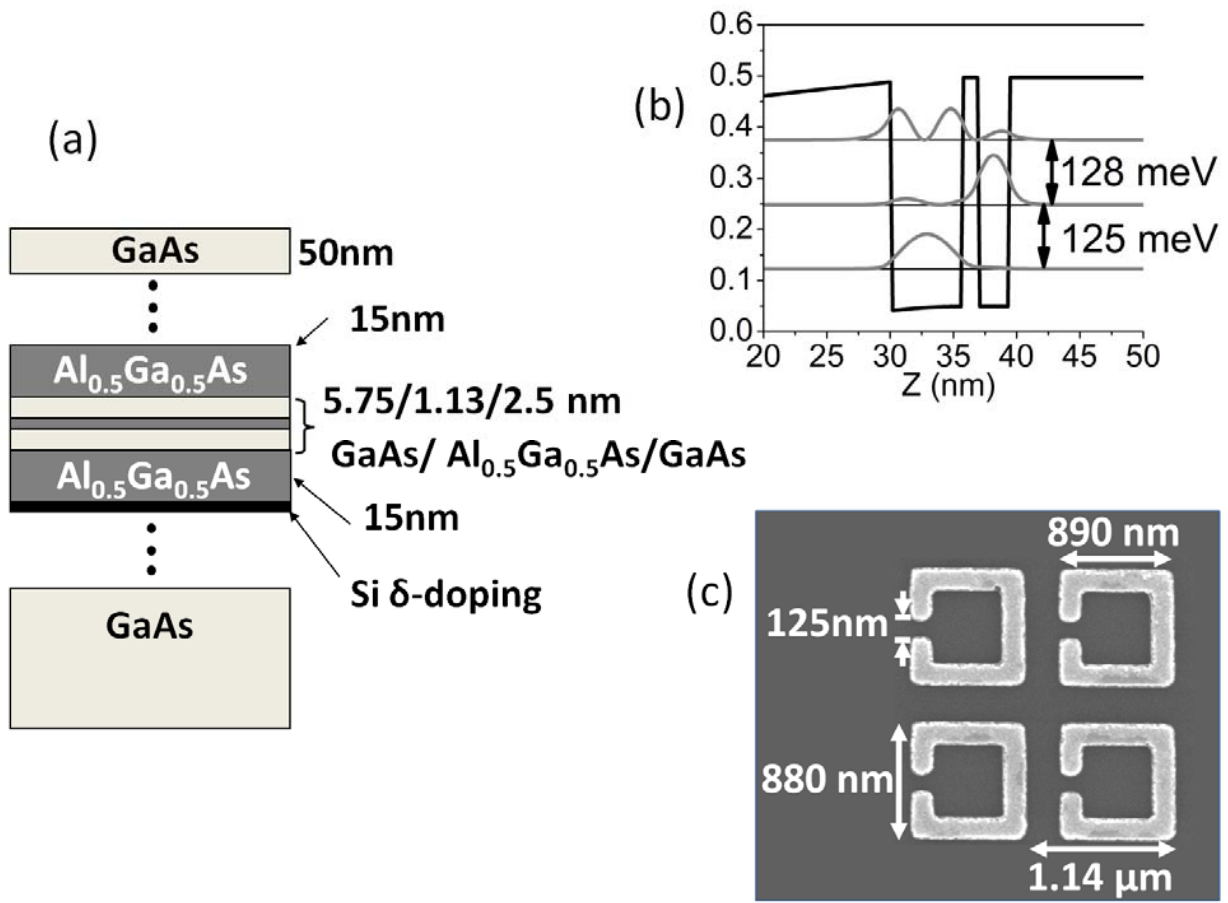

Fig. 2: (a) - The layer sequence of the QW sample. (b) An energy band diagram of two asymmetric coupled QWs. The subbands and their corresponding wavefunctions (modulus) are shown as well. (c) A SEM image of representative SRRs.

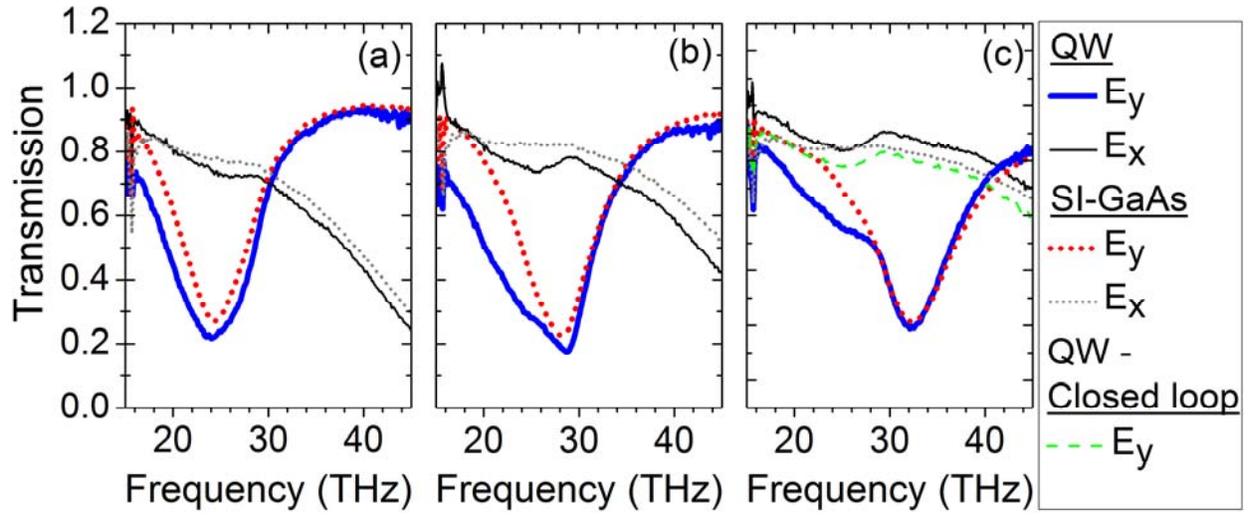

Fig. 3: (a-c) Transmission spectra for three different SRR arrays, which were scaled (length, linewidth, gap size and periodicity) to resonate at different frequencies. (a-c) are scaled by 0.9, 1.0 and 1.1, respectively. The legend indicates the polarization of the incident light for each transmission line and the substrate beneath the SRRs.

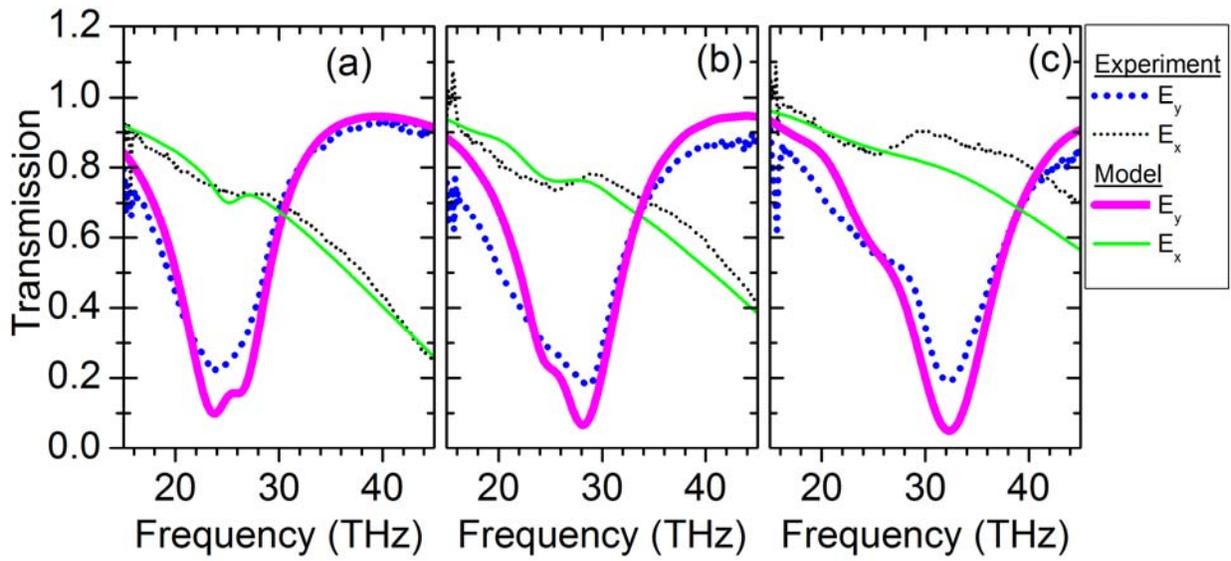

Fig. 4: (a-c) Experimental and calculated transmission spectra for three different SRR arrays located on the QW sample, which were scaled (length, linewidth, gap size and periodicity) to resonate at different frequencies. The legend indicates the polarization of the incident light for each transmission.